\documentclass[aip,amsmath,amssymb,reprint]{revtex4-2}
\pdfoutput=1

\usepackage{silence}
\usepackage[utf8]{inputenc}
\usepackage{graphicx}
\usepackage{tabularx}
\graphicspath{{Figures/}}
\usepackage{xcolor}
\usepackage{dcolumn}
\usepackage{bm}
\usepackage[breaklinks]{hyperref}
\usepackage[babel=true]{csquotes}

\usepackage[T1]{fontenc}
\usepackage{mathptmx}
\usepackage{etoolbox}
\usepackage{makecell}
\usepackage{comment}

\usepackage{amsmath}

\usepackage{siunitx}
\usepackage{caption}
\DeclareSIUnit{\fps}{ \translate{frames per second} }

\newcommand{\tate}{\texorpdfstring{1\textit{T$\,'$}-TaTe\textsubscript{2}}{1\textit{T}-TaTe2}}
\newcommand{\tas}{\texorpdfstring{1\textit{T}-TaS\textsubscript{2}}{1\textit{T}-TaS2}}
\newcommand{\tase}{\texorpdfstring{1\textit{T}-TaSe\textsubscript{2}}{1\textit{T}-TaSe2}}

\usepackage{float}
\newcommand{\uproman}[1]{\uppercase\expandafter{\romannumeral#1}}

\begin{document}

\title{Megahertz cycling of ultrafast structural dynamics enabled by nanosecond thermal dissipation}

\author{Till Domröse}
\author{Leonardo da Camara Silva}
\author{Claus Ropers}
\email[Corresponding author: ]{claus.ropers@mpinat.mpg.de}
\affiliation{Department of Ultrafast Dynamics, Max Planck Institute for Multidisciplinary Sciences, 37077 Göttingen, Germany}
\affiliation{4th Physical Institute -- Solids and Nanostructures, University of Göttingen, 37077 Göttingen, Germany}

\begin{abstract}
Light-matter interactions are of fundamental scientific and technological interest. Ultrafast electron microscopy and diffraction with combined femtosecond-nanometer resolution elucidate the laser-induced dynamics in structurally heterogeneous systems. These measurements, however, remain challenging due to the brightness limitation of pulsed electron sources, leading to an experimental trade-off between resolution and contrast. Larger signals can most directly be obtained by higher repetition rates, which, however, are typically limited to a few \SI{}{\kHz} by the thermal relaxation of thin material films. Here, we combine nanometric electron-beam probing with sample support structures tailored to facilitate rapid specimen cooling. Optical cycling of a charge-density wave transformation enables quantifying the mean temperature increase induced by pulsed laser illumination. Varying the excitation fluence and repetition rate, we gauge the impact of excitation confinement and efficient dissipation on the heat diffusion in different sample designs. In particular, a thermally optimized support can dissipate average laser intensities of up to \SI{200}{\micro\watt\per\um^2} within a few nanoseconds, allowing for reversible driving and probing of the CDW transition at a repetition rate of \SI{2}{\MHz}. Sample designs tailored to ultrafast measurement schemes will thus extend the capabilities of electron diffraction and microscopy, enabling high-resolution investigations of structural dynamics.
\end{abstract}

\maketitle

Electron microscopy and diffraction are versatile experimental tools used to explore steady-state dynamics and image heterogeneous systems in both materials and life science. Over the past years, advances in aberration correction \cite{haider1998}, detector development \cite{Levin2021} and sample preparation \cite{mayer2007,russo2014} have enabled profound insights in the fields of structural biology \cite{cheng2015,yip2020}, heterogeneous catalysis \cite{chee2023} and solid-state physics \cite{scott2012,Ophus2019,hawkes2019} with down to atomic resolution. Beyond the imaging of stationary structures, ultrafast electron microscopy and diffraction promise the investigation of non-equilibrium processes and transient states of matter\cite{alcorn2023,filippetto2022,lee2024}. Such dynamics are induced by ultrashort optical or electrical pulses and evolve on femtosecond to picosecond time scales, far below the shutter speeds of the fastest available electron detectors. Pioneering works in ultrafast transmission electron microscopy \cite{domer2003,zewail2010} have allowed to directly image phenomena such as structural phase transformations \cite{VanderVeen2013,Danz2021b} and strain-wave propagation \cite{Feist2017,mckenna2017,nakamura2020} on nanometer length scales. Similarly, in ultrafast electron diffraction (UED) \cite{siwick2003}, atomic-scale information on evolving crystalline order and phonon populations are obtained by tracing diffraction intensities \cite{siwick2003,ungeheuer2024,Eichberger2010,erasmus2012,morrison2014,Han2015,Haupt2016,Sie2019,horstmann2020,Kogar2020,ji2020,cheng2022,domrose2024}, spot profiles \cite{Vogelgesang2018,Storeck2021,domrose2023,cheng2024}, and momentum-dependent diffuse scattering \cite{Waldecker2017,Otto2021,kurtz2024}.

\begin{figure*}[ht]
\centering
\includegraphics[scale=1]{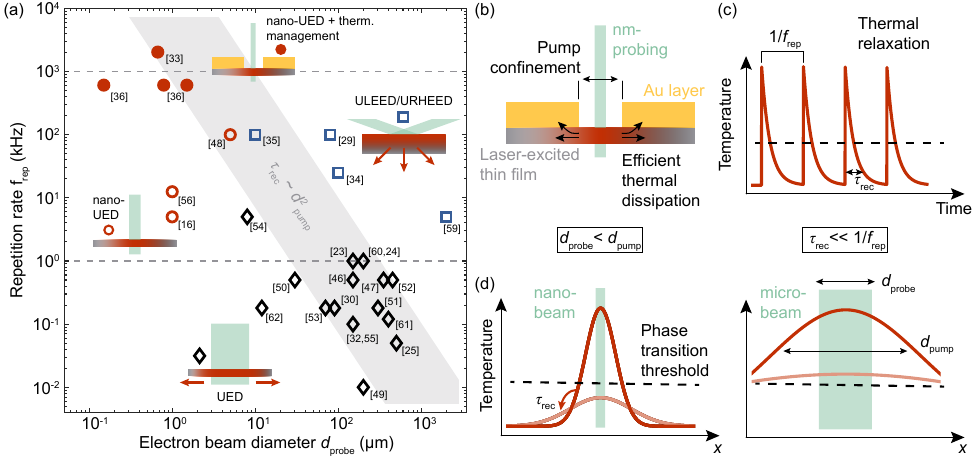}
\caption{
\textbf{Ultrafast electron diffraction of structural phase transformations.}
\textbf{a} Literature overview of repetition rates $f_\text{rep}$ and collimated electron beam diameters in UED probing structural phase transitions. In-plane thermal dissipation in thin material films limits the repetition rates for \num{100}\textendash\SI{}{\um} probing (black diamonds) to \SI{1}{\kHz}. Surface-sensitive techniques (blue squares) offer faster thermal relaxation into the bulk, and thus higher duty cycles. Smaller pump beams allow to enhance $f_\text{rep}$ also in transmission probing (red open circles). Combined nanobeam diffraction and optimized thermal management (filled red circles) extend the available $f_\text{rep}$ into the \SI{}{\MHz}-regime.
\textbf{b} Schematic sample design in nano-UED. A gold reflection layer confines the optical excitation, and ensures efficient thermal dissipation.
\textbf{c} Stroboscopic probing requires sufficient thermal relaxation between subsequent excitation events.
\textbf{d} Small probe volumes allow for smaller pump profiles (diameter $d_\text{pump}$), leading to faster sample cooling times $\tau_\text{rec}$ and higher higher $f_\text{rep}$.
}
\label{fig:PumpParameters}
\end{figure*}

These measurements rely on the stroboscopic principle: femtosecond probe pulses take snapshots of the non-equilibrium state at a temporal delay $\Delta t$ after the optical excitation \cite{basov2017,buzzi2018,delatorre2021,filippetto2022,lee2024}. Critically, such a scheme is only sensitive to reversible processes, as sufficient contrast in imaging or diffraction arises from averaging over many individual probe events at a set pump-probe delay. A full reconstruction of the structural dynamics thus necessitates complete relaxation within one pump-probe cycle. This entails sample cooling prior to the arrival of the subsequent laser excitation \cite{kazenwadel2023}, as well as, in the case of structural phase transformations, a re-formation of the initial phase (Fig.~\ref{fig:PumpParameters}c). For a given total measurement time, the maximum available repetition rate $f_\text{rep}$ satisfying these conditions will determine the signal-to-noise ratio of the experiment.

In studying laser-induced phase transitions, often, a certain threshold fluence is required. At such a given energy density in an individual excitation pulse, cumulative heating, scaling with the overall absorbed laser power \cite{lax1977}, will increase linearly with the repetition rate. Thus, obtaining enhanced contrast by higher duty cycles may be realized in two ways. First, cooling down to a lower initial temperature compensates for some additional average temperature increase. Second, the thermal relaxation time in thin-film specimens, typically investigated in transmission diffraction and imaging, scales quadratically with the pump beam diameter (Fig.~\ref{fig:PumpParameters}d; and Supplementary Material). Consequently, reducing the laser-excited area facilitates reversible high-repetition rate driving also in materials with reduced thermal conductivity, and phase diagrams characterized by a succession of multiple structural orders within a small temperature range become fully accessible to ultrafast methodology.

The subtle, sub-\r{A}ngstrom structural changes associated with phase transformations are resolved by ultrafast electron diffraction \cite{sciaini2009,Eichberger2010,erasmus2012,VanderVeen2013,morrison2014,Han2015,Haupt2016,Sie2019,he2016,sun2015,li2016,tao2016,LeGuyader2017,ji2020,Kogar2020,Storeck2021,siddiqui2021,sood2021,li2022,cheng2022,xu2023,shiratori2024,domrose2023,siddiqui2023,diaz2024,domrose2024,wall2012,yang2016,Vogelgesang2018,horstmann2020,weathersby2015,shen2018}. As a prerequisite for spatially-averaging measurements, a clear interpretation of the data requires a homogeneous excitation across the electron beam diameter $d_\text{probe}$, limiting the smallest possible focus of the laser spot. Figure~\ref{fig:PumpParameters}a displays documented experimental parameters in prior UED characterizations of structural transitions. Even though these investigations span a broad range of systems and beam energies, the quadratic relationship between cooling time and pump beam diameter translates into a fundamental limitation of the available repetition rate. Importantly, surface-sensitive techniques allow for higher duty cycles \cite{wall2012,yang2016,Vogelgesang2018,horstmann2020}, as thicker samples additionally feature thermal dissipation into the bulk.

Electron beam diameter and divergence, which directly affects the momentum resolution in diffraction, are defined by the emittance of the beam used. A small effective electron source size is thus beneficial for both, high repetition-rate pumping and high-resolution probing. Setups employing micrometer-sized, flat photocathodes operate in the nanometer-radian emittance regime \footnote{Alternatively, emittances may be specified in \SI{}{\um\milli\radian}}, corresponding to electron beam diameters of tens to hundreds of micrometers. Dynamics are excited and probed at rates between \SI{100}{\Hz} and \SI{1}{\kHz}. In contrast, picometer-radian beam emittances offer enhanced resolution in collimated electron nanobeams, reduced spatial averaging, confined excitation, and faster sample cooling times.

In this work, we investigate the thermal properties of tailored sample environments optimized for rapid thermal relaxation. A confinement of the laser excitation and enhanced thermal dissipation enable nanosecond cooling times, allowing for megahertz cycling of structural dynamics and transitions in thin material films. We benchmark these capabilities in a nanobeam ultrafast electron diffraction (nano-UED) study of a structural phase transformation in the layered material \tate{} \cite{siddiqui2021,domrose2024}. Structural dynamics are reversibly driven at a repetition rate of \SI{2}{\MHz} and an incident fluence of \SI{10}{\mJ\per\cm^2}, and probed by a high-coherence pulsed electron nanobeam. Featuring also high momentum resolution, our measurements advance ultrafast electron diffraction into the parameter space of picometer-radian beam emittances, nanometer beam diameters, quantified femtosecond pulse durations, and megahertz repetition rates.

\begin{figure*}[htp]
\centering
\includegraphics[scale=1]{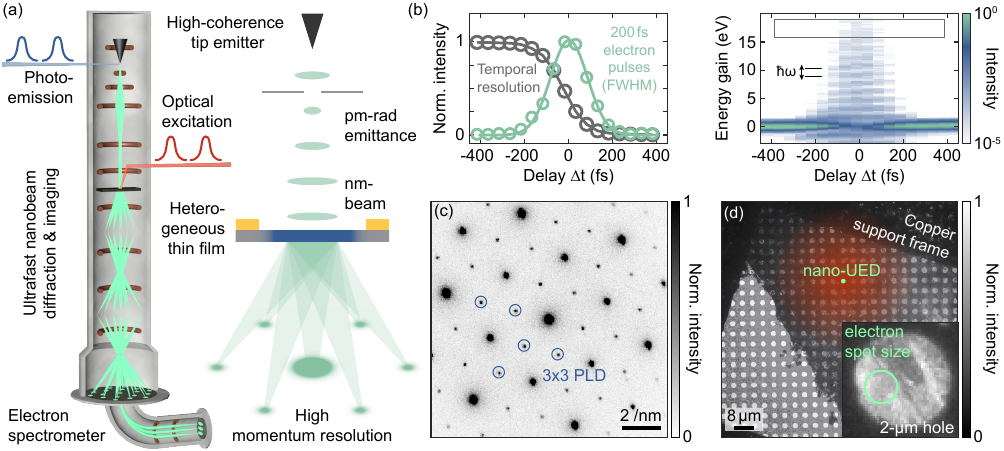}
\caption{
\textbf{Nanobeam ultrafast electron diffraction.}
\textbf{a} Ultrafast transmission electron microscope. Electron pulses are generated via photoemission from a tip emitter. Laser-induced dynamics in heterogeneous thin films are probed at high \textit{k}-space resolution with a collimated nanobeam, enabled by a picometer-radian transverse beam emittance.
\textbf{b} \textit{In-situ} pulse-duration measurement in the sample plane (right). Inelastic electron-light scattering leads to the population of discrete side-bands in the electron spectra separated by the photon energy $\hbar\omega$. The intensity in the outer energy sidebands (grey rectangle, integrated along the energy axis, cumulatively summed along the time axis) yields a quantitative measure of the instrument resolution function (left, grey). Its derivative corresponds to the electron pulse structure (green). We obtain durations of \SI{200}{\fs}.
\textbf{c} Example diffractogram of the $3\times3$ phase of \tate{} measured before time zero, recorded with ultrashort electron pulses (beam diameter \SI{670}{\nm}) and under full thermal load (\SI{10}{\mJ\per\cm^2} incident fluence at \SI{2}{\MHz} rate). The superstructure formation leads to low-intensity second-order satellites (blue circles), the order parameter of the structural transformation.
\textbf{d} Electron micrograph of a \tate{} thin film suspended below a gold aperture array (\SI{2}{\um} hole diameter, \SI{4}{\um} pitch). Inset: enlarged view of a single \num{2}\textendash\SI{}{\um} hole. The green circle illustrates the electron spot size.
}
\label{fig:Setup}
\end{figure*}

Our experiments are carried out in an ultrafast transmission electron microscope (UTEM) \cite{zewail2010,piazza2013,Feist2017,Cremons2017,houdellier2018,zhu2020,olshin2020,kuttruff2024,weber2024,fu2020,ji2020,dahan2021,kim2023,alcorn2023}. Non-equilibrium dynamics are excited by ultrashort laser pulses with tunable wavelength and repetition rate [here: \SI{800}{\nm} wavelength, \SI{50}{\fs} duration, between \SI{101}{\kHz} and \SI{2}{\MHz} repetition rate; Fig.~\ref{fig:Setup}(a)], and probed by ultrashort electron pulses generated via linear photoemission. The Göttingen UTEM features a high-brightness pulsed field-emitter electron source (\num{120}\textendash\SI{200}{\keV} electron energy) \cite{Feist2017}. The confinement of the photoemission to the front-apex of the emitter tip yields a particularly small effective source size, and thus enables nanoscale investigations of, e.g., charge-density wave (CDW) transformations at high momentum resolution \cite{Vogelgesang2018,Storeck2021,Danz2021b,domrose2023,domrose2024}. While such electron sources are primarily found in UTEMs \cite{Feist2017,houdellier2018,zhu2020,olshin2020,kuttruff2024,weber2024}, their usage is not restricted to electron microscopes, and has also been demonstrated in dedicated UED setups for investigations of structural dynamics in transmission \cite{gulde2014} and reflection \cite{Storeck2021}.

The versatility of electron microscopy in forming electron beams enables additional control over other key experimental parameters such as the temporal resolution. In particular, spectroscopic characterizations of inelastic electron scattering at optical near-fields [photon-induced near-field electron microscopy, PINEM \cite{barwick2009,Feist2015a,dahan2021}; Fig.~\ref{fig:Setup}(b)] allow for a quantitative determination of the electron pulse duration and shape, as well as time zero \cite{plemmons2014}. For low pulse charges, we obtain a pulse duration of \SI{200}{\fs} (full-width-at-half maximum). Importantly, this pulse characterization can be performed \textit{in-situ}, accounting for small changes of the sample position and tilt that may alter the relative timing between pump and probe pulses. Measured signals can then be corrected by an instrument resolution function recorded under the same experimental conditions \cite{domrose2024}. Direct access to the electron pulse properties also allows for balancing the bunch charge and temporal resolution, optimizing image contrast at the expense of moderate Coulomb-induced temporal pulse broadening \cite{siwick2002,wang2009,aidelsburger2010,haindl2023}, tailored to the fastest features in the investigated dynamics.

\begin{figure}[ht!]
\centering
\includegraphics[scale=1]{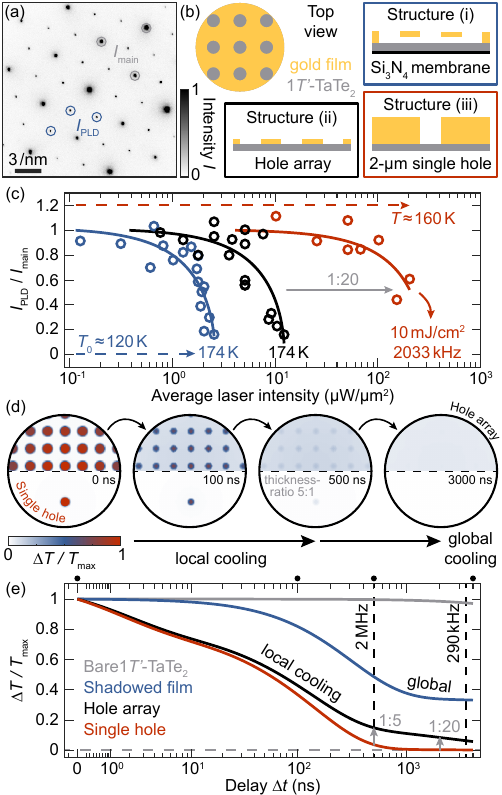}
\caption{
\textbf{Thermal characterization of sample supports.}
(a)~Example diffractogram of \tate{}. The intensity of $(3\times3)$ PLD spots (red circles) measured with a continuous-wave electron beam is indicative of cumulative sample heating.
(b)~Schematics of the sample supports. Arrays of micrometer-sized gold apertures on top of a \tate{} thin film confine the optical excitation [structure (i)], and additionally provide an efficient heat bath [structure (ii)]. A single-hole design [structure (iii)] enables optimized thermal dissipation.
(c)~Normalized PLD spot intensities under high thermal loads and employing the support structures in (b). The solid lines are phenomenological fits to estimate the average temperature rise (see Supplementary Material).
(d)~Simulated temperature profiles in a hole array (top half) and in the single-hole support (bottom). After laser excitation, local heat dissipation around the temperature hotspots is followed by global cooling where the temperature of the surrounding bath depends on the overall pump volume determined by the gold filling factor.
(e)~Temporal evolution of the average temperature in the central $\num{2}$\textendash$\mu\text{m}$ sample region derived from the simulation. The thicker gold film in structure (iii) reduces the temperature after \SI{500}{\ns} by a factor of five compared to structure (ii), while the reduced pump volume leads to a \num{20}-fold temperature reduction at \SI{2}{\us}, in agreement with the experimental data displayed in (c).
}
\label{fig:SampleCharacterization}
\end{figure}

In the following, we illustrate the technical capabilities of the setup by investigating charge-density wave dynamics in the layered material \tate{}. Comprised out of weakly-bound tellurium-tantalum-tellurium trilayers, the reduced dimensionality and strong electron-phonon coupling favour the formation of charge-density waves (CDWs) accompanied by a large-amplitude periodic lattice distortion (PLD) \cite{sorgel2006,elbaggari2020}. These types of charge-ordered phases occur in a range of quantum materials \cite{pouget2024}, and were investigated by ultrafast diffraction and microscopy employing x-rays \cite{beaud2014,gerber2017,Laulhe2017,singer2018,johnson2024} or electrons \cite{Eichberger2010,erasmus2012,morrison2014,Han2015,Haupt2016,Sie2019,horstmann2020,Kogar2020,ji2020,domrose2024,domrose2023,Danz2021b,Vogelgesang2018,Storeck2021,cheng2024,kurtz2024,sun2015,li2016,tao2016,LeGuyader2017,siddiqui2021,sood2021,yang2016,wall2012,cheng2022}, including investigations of the structural transitions in \tate{} \cite{siddiqui2021,domrose2024}. The material's room-temperature phase is characterized by a $(3\times1)$ superstructure when compared to the undistorted 1$T$-polytype found in, e.g., the chemically related compounds \tas{} and \tase{}. Below a temperature of \SI{174}{\kelvin}, the system undergoes a first-order structural transformation \cite{sorgel2006}. In electron diffractograms, the appearance of this $(3\times3)$ phase is identified by additional low-intensity satellite peaks surrounding the high-intensity reflections of the undistorted host lattice [Fig.~\ref{fig:SampleCharacterization}(a)]. 

The first-order transition allows for electron diffraction measurements of the time-averaged temperature increase under various incident laser fluences and repetition rates. Specifically, the cumulative thermal response is encoded in the intensity of the second-order $(3\times3)$ spots, i.e., the order parameter of the transformation both in- and out-of equilibrium, when measured by a continuous electron beam (Fig.~\ref{fig:SampleCharacterization}). For sufficiently low laser intensities, and starting from an estimated base temperature of \SI{120}{\kelvin}, the presence of PLD spots in the diffractograms indicates that the sample remains in its low-temperature state for the majority of the pump cycle. Steady-state heating at higher thermal loads increases the mean temperature, and a suppression of the PLD amplitude sets in upon approaching the phase transition threshold. Finally, for an insufficient temporal separation of the excitation pulses, the average temperature rises above \SI{174}{\kelvin}, and the PLD spots disappear. As expected for cumulative heating, we find that different combinations of repetition rates and fluences that correspond to the same average intensity yield the same thermal suppression (see also Supplementary Fig.~S1).

We measure this temporal average over the laser-induced dynamics in sample support structures comprised of different arrangements of circular apertures in a gold film [Fig.~\ref{fig:SampleCharacterization}(b)], showing rather different thermal responses. This includes (i)~hole arrays placed above a \tate{} sample after deposition on a SiN membrane, (ii) arrays directly suspending a \tate{} flake, as well as (iii) a \tate{} film spanned across a single aperture.

The gold film reflects a substantial amount of the incident laser radiation (see Supplementary Material), minimizing the absorbed laser power and, thus, the steady-state temperature increase \cite{lax1977}. In structure (i), a square array of apertures with a diameter and separation of \SI{2}{\um} in the $\num{50}$\textendash$\SI{}{\nm}$ gold layer \cite{russo2014} shields about \SI{80}{\percent} of the \tate{} film compared to a bare \tate{} sample and confines the absorbed excitation to local hotspots. The coherence of the employed electron source allows for scaling down the probe beam diameter below this engineered excitation heterogeneity, while maintaining high transverse momentum resolution \cite{domrose2023}. In such nano-UED measurements, the highest possible laser intensity will only be limited by the average heating within the central \num{2}\textendash\SI{}{\um} hole.

Heat transport simulations quantitatively capture the spatiotemporal relaxation after an instantaneous temperature increase [Fig.~\ref{fig:SampleCharacterization}(d), (e); see also Supplementary Material and Fig.~S1]. In order to gauge the suitability of the support structures for nano-UED measurements across a larger range of materials and systems, we focus on the behavior on time scales comparable to the pulse-to-pulse separation set by the repetition rate. The segmentation of the excitation into an array of smaller pump spots [Fig.~\ref{fig:SampleCharacterization}(d)] leads to a pronounced cooling already within a microsecond, which represents a drastic enhancement compared to an unobstructed absorption [blue and grey curve in Fig.~\ref{fig:SampleCharacterization}(e), respectively]. After the local heat bath is exhausted, i.e., when the temperatures in shadowed and unshadowed regions have equalized, the macroscopic dissipation evolves as in the case of an unstructured excitation, but with substantially reduced thermal load. Nevertheless, as the relaxation is governed by the comparably low thermal conductivity of the \tate{} film, cumulative heating suppresses the PLD spots already at intensities of around \SI{2}{\micro\watt\per\um^2} (here: \SI{0.8}{\mJ\per\cm^2} at \SI{254}{\kHz}), at which the average temperature increases by more than \SI{50}{\kelvin} [blue curve in Fig.~\ref{fig:SampleCharacterization}(c)]. We envision such a design to be used primarily in special cases where delicate specimens such as few-layer heterostructures require a continuous support membrane.

\begin{figure}[ht]
\centering
\includegraphics[scale=1]{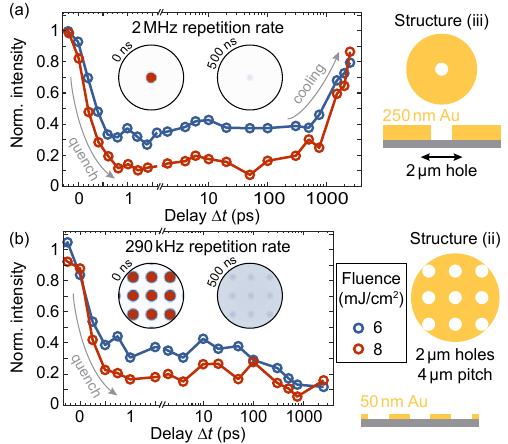}
\caption{
\textbf{Nano-UED measurements at high repetition rates.}
(a) Delay curves for second-order $(3\times3)$ PLD spot intensities after the optical excitation in support structure (iii) \cite{domrose2024}. The transition is reversibly cycled at a repetition rate of \SI{2}{\MHz}. Enhanced thermal dissipation leads to a recovery of the low-temperature phase after \SI{4}{\ns}. Insets: simulated temperature profiles as displayed in Figure~\ref{fig:SampleCharacterization}.
(b) Delay curves similar to (a), recorded at a repetition rate of \SI{290}{\kHz} and with a \tate{} thin film in support structure (ii). While the PLD is reestablished before the arrival of the next pump pulse, cooling times are longer than for the single-hole sample.
}
\label{fig:TaTe2_ultrafast}
\end{figure}

Placing the millimeter-sized \tate{} flake directly behind and in contact with the aperture array [structure (ii)] further reduces the average heating by about a factor of five [black curve in Fig.~\ref{fig:SampleCharacterization}(c) and (e)]. As is evident from the heat diffusion simulations, and compared to a free-standing film, both the high thermal diffusivity of the surrounding gold and the additionally available heat bath accelerate the specimen cooling. These results illustrate the improvements gained in ultrafast diffraction experiments that combine a reduced excitation volume with efficient thermal dissipation. For \num{2}\textendash\SI{}{\um} holes, the local dynamics are largely completed after around \SI{500}{\ns}. Accordingly, an efficient management of the heat left to dissipate globally promises reversible cycling at hundreds of kilohertz to megahertz rates, depending on the number of excited holes, their diameters and distances, and the gold film thickness.

Consistent with an estimation based on these different influences on the thermal properties of the support structures (see Supplementary Material), an individual aperture and a \num{250}-nm thick gold film achieve a reduction in average heating by more than two orders of magnitude [red curves in Fig.~\ref{fig:SampleCharacterization}(c)], whereby the difference in gold film thickness contributes the expected factor of five already after the local relaxation has completed [Fig.~\ref{fig:SampleCharacterization}(e)]. As we show here, such a sample [support (iii)] can accommodate laser intensities of up to \SI{200}{\micro\watt\per\um^2}, at which we observe minor cumulative heating, and an estimated temperature increase of around \SI{40}{\kelvin} (see Supplementary Material). In comparison to support (ii), these values correspond to a \num{20}-fold increase of the usable average laser intensity.

All three investigated sample supports enable ultrafast electron diffraction measurements across a broad fluence range and at very high repetition rates, whereby the highest possible duty cycles far exceed values achieved by placing the sample close to a copper grid bar or the edge of a silicon wafer [Fig.~\ref{fig:PumpParameters}(a)]. In agreement with the simulation results, support structure (ii) allows for stroboscopic probing at \SI{290}{\kHz} [Fig.~\ref{fig:TaTe2_ultrafast}(b)]. Furthermore, we employed a single-aperture sample [structure (iii)] in the investigation of CDW transformations in \tas{} via imaging \cite{Danz2021b} and diffraction \cite{domrose2023} at up to \SI{609}{\kHz} repetition rate, limited by the maximum available pulse rate of the laser setup used. Investigating the structural transitions in \tate{}, we now enter UED into the \SI{}{\MHz}-regime [Fig.~\ref{fig:TaTe2_ultrafast}(a)] \cite{domrose2024}. In these measurements, the second-order PLD spots in the diffractograms recorded under highest thermal load [\SI{10}{\mJ\per\cm^2} and \SI{2}{\MHz} in Fig.~\ref{fig:Setup}(c)] are both clearly visible and sharp, underlining the drastically reduced cumulative heating and the high reciprocal-space resolution for the \num{670}\textendash\SI{}{\nm} collimated electron illumination.

The structural dynamics are evident from the temporal evolution of the $(3\times3)$ diffraction spot intensities which can be quantitatively mapped onto the PLD amplitude by dynamical diffraction simulations \cite{domrose2024}. Higher repetition rates allow us to reduce the image acquisition times without compromising the temporal resolution or the signal-to-noise ratio, enabling the recording of detailed datasets that cover a large parameter range, and which are not influenced by long-term sample drifts. For the dynamics excited at \SI{2}{\MHz} [Fig.~\ref{fig:TaTe2_ultrafast}(a)], we integrate for a period of \SI{90}{\s} per set time delay, and, in conjunction with direct electron detection, derive a signal-to-noise ratio of \num{50}. In comparison, a noise level of \SI{5}{\percent} is achieved in the measurements with support structure~(ii) at \SI{290}{\kHz} for image acquisition times of \SI{4}{\minute} per delay [Fig.~\ref{fig:TaTe2_ultrafast}(b)]. At early delays, both experiments show the temporal PLD amplitude evolution reported previously for \tate{} \cite{domrose2024} that is in line with the dynamics in other TMDC-CDW phases \cite{Hellmann2012b}. The structural distortions are suppressed within \SI{500}{\fs}, followed by a partial recovery for low and intermediate pump fluences. Presumably, the slightly larger variations in the recorded spot intensities in the second set of measurements are related to the fragility of the \num{50}\textendash\SI{}{\nm} gold net that carries the specimen. The optical excitation induces a global oscillation of the sample that modulates all scattered intensities at larger delays, inducing a second PLD spot suppression beyond \SI{10}{\ps}. Such effects could be reduced by thicker gold films, increasing the rigidity of the support. Furthermore, the thermal recovery of the low-temperature phase occurs at temporal delays outside the measurement range but is sufficiently fast to reestablish the $(3\times3)$ PLD before the arrival of the next pump pulse after \SI{3.5}{\us}. In comparison, long-term oscillations are absent for the single-hole structure in the \num{2}\textendash\SI{}{\MHz} measurements, and the PLD amplitude reverts to its initial configuration after only \SI{4}{\ns}. Therefore, even higher thermal loads might be compatible with the sample design, particularly when combined with a larger temperature difference between the initial and the final structural state.

In conclusion, electron-source coherence and thermal dissipation limit resolution and contrast in ultrafast diffraction experiments. Both small effective electron source sizes and sample designs tailored to high-repetition-rate driving are required to overcome these experimental challenges, enabling the formation of collimated electron nanobeams with high transverse momentum and temporal resolution that probe dynamics at megahertz repetition rates. An immediate benefit of smaller probe volumes and high-contrast data acquisition is the possibility of avoiding micrometer spatial inhomogeneities contributing to the recorded dynamics. This includes varying crystal orientations, the influence of sample edges, and heterogeneous excitation densities. Moreover, nanoscale probe beams will enable the investigation of heterostructures comprised of different compounds, and functional devices that exploit the tunability of quantum materials \cite{mak2016}. For the latter, electrical contacting of thin films includes patterning with thick metallic layers that, simultaneously, can serve as laser reflection layers and efficient thermal conductors, analogous to the gold apertures in the support structures presented here. As a consequence, we expect that high repetition rates are also applicable to these types of samples.

Additionally, due to the strong interaction of electron beams with matter, quantitative data evaluations either require explicitly accounting for multiple scattering, or considering integrated diffraction spot intensities, i.e., averaging over the rocking curve of a reflection. The latter is almost always given in large specimens, as even single-crystal samples feature a locally varying morphology. Therefore, while qualitative evaluations of dynamics in different Laue zones are possible \cite{LeGuyader2017,cheng2022}, quantifying structural dynamics parallel to the incident electron illumination seems challenging for \num{100}\textendash\SI{}{\um} beams. In contrast, reducing the spatial averaging while correlating the measurements to dynamical diffraction simulations immediately enables recording and analyzing rocking curve dynamics \cite{domrose2023}.

Alternatively, such tilt-series ultrafast diffraction may be used to map the entire reciprocal lattice by combining beam- and sample-tilting as employed in, e.g., precession electron diffraction \cite{gemmi2019}, which was recently transferred to the ultrafast time domain \cite{shiratori2024}. In the future, these measurements will allow for both a complete structural refinement with femtosecond resolution and the extraction of element-specific dynamics. Such characterizations may be particularly beneficial for investigations of materials with more complex unit cells \cite{mankowsky2015}.

Beyond ultrafast diffraction measurements, the increase in coherent current gained by higher repetition rates can also directly compensate for a reduction of the electron probe signal resulting from smaller beam-limiting apertures. Approaching an ideal, fully-coherent electron point source, this type of beam shaping is routinely found in continuous-wave TEM techniques that operate at higher electron emission from the source \cite{hawkes2019}. As such beam-shaping leaves the transverse coherence length unchanged, sample designs as presented here seem imperative to extend the possibilities of UTEM to reach combined atomic-scale spatial and femtosecond temporal resolutions. Finally, the drastically accelerated sample cooling after optical excitation may enhance the observation of transient states in biological specimens by cryo-electron microscopy, offering a more rapid, nanosecond or even faster revitrification of samples after laser-induced ultrafast melting \cite{harder2022}.

\section*{Supplementary Material}
The Supplementary Material contains further information on the experimental setup and the sample preparation, and descriptions of the thermal transport simulations and the experimental characterizations of cumulative heating in the different sample supports.

\section*{Acknowledgments}
The authors thank M.~Sivis for technical support in focused ion beam milling. Furthermore, we gratefully acknowledge insightful discussions with I.~Vinograd, and continued support from the Göttingen UTEM team.
This work was funded by the Deutsche Forschungsgemeinschaft (DFG, German Research Foundation) in the Collaborative Research Centre ``Atomic scale control of energy conversion'' (217133147/SFB 1073, Project No. A05) and via resources from the Gottfried Wilhelm Leibniz Prize (RO 3936/4-1).

\section*{Author contributions}

T.D. and L.d.C.S. conducted the experiments and simulations, analyzed the data, and prepared the specimens. C.R. directed the study. All authors discussed the results and their interpretations. T.D. and C.R. wrote the manuscript with discussions and input from all authors.

\section*{Competing interests}

T.D. and C.R. contributed to Ultrafast Electron Microscopy technology used in this work, developed at the University of Göttingen and the Max Planck Institute for Multidisciplinary Sciences, and subsequently licensed to JEOL-IDES in 2024.

\section*{Data availability}

The data that support the findings of this study are available from the corresponding author upon reasonable request.

\bibliography{TaTe2}

\end{document}


\title{\textmd{Supplementary Material for\\} Megahertz cycling of ultrafast structural dynamics enabled by nanosecond thermal dissipation}

\author{Till Domröse}
\author{Leonardo da Camara Silva}
\author{Claus Ropers}
\email[Corresponding author: ]{claus.ropers@mpinat.mpg.de}
\affiliation{Department of Ultrafast Dynamics, Max Planck Institute for Multidisciplinary Sciences, 37077 Göttingen, Germany}
\affiliation{4th Physical Institute -- Solids and Nanostructures, University of Göttingen, 37077 Göttingen, Germany}

\maketitle

\section{Setup and data acquisitions}
The Göttingen UTEM is a modified \enquote{JEOL JEM-2100F}, equipped with a Schottky-type ZrO/W emitter. Our laser setup is a \enquote{Light Conversion Carbide} (\SI{40}{\fs} pulse duration, \SI{1030}{\nm} wavelength, up to \SI{2}{\MHz} repetition rate) whose output is split into two optical paths. A fraction is frequency-doubled and used for the generation of ultrashort electron pulses via linear photoemission by illuminating the electron emitter. The remaining laser light is coupled into an optical parametric amplifier (\enquote{Light Conversion Orpheus F}) where the laser pulses are converted to \SI{800}{\nm} wave length and coupled into the microscope's column, illuminating the sample under close to perpendicular incidence. The laser focus in the sample plane amounts to \SI{40}{\um} full-width at half-maximum.
Electron diffractograms are recorded with a direct electron detector (\enquote{Direct Electron DE16}) in electron-counting mode. The thermal characterizations of the different support structures include image acquisition times of \SI{100}{\s} per measured laser intensity. For the time-resolved data, we integrate for a total of \SI{90}{\s} per delay in the \SI{2}{\MHz} measurements, and for \SI{4}{\minute} in the measurements conducted at \SI{290}{\kHz}.
The spectroscopic electron pulse characterizations were conducted using an electron spectrometer and energy filtering device (\enquote{CEOS CEFID}), and spectrograms were recorded with a hybrid pixel detector based on the Timepix3 chip (\enquote{Amsterdam Scientific Instrument's Cheetah T3}). The TEM sample holder for cooling with liquid nitrogen was a \enquote{Gatan Liquid Nitrogen Cooling Holder, model \num{636}}. The estimated base temperature is \SI{120}{\kelvin}.

\section{Sample preparation}

Thin film \tate{} crystals (\enquote{HQ graphene}) were cut to a nominal thickness of \SI{50}{\nm} by Ultramicrotomy.
For sample design (i), the \tate{} flakes were transferred onto a \SI{50}{\nm} silicon nitride membrane spanned over a standard \SI{3}{\mm} silicon wafer. The gold aperture arrays used for support structures (i) and (ii) are commercially available TEM grids (\enquote{Quantifoil UltrAuFoil R2/2}) with \SI{2}{\um}-sized holes and \SI{4}{\um} pitch \cite{russo2014}. The gold film has a thickness of \SI{50}{\nm}. Placing the gold masks over the silicon wafer with micrometer precision is realized with a custom-made apparatus, and the arrays are glued to the silicon wafers with poly(methyl 2-methylpropenoate).
Sample support (ii) features the same TEM grids, but the \tate{} specimen was suspended onto the sample carrier directly.
Finally, support (iii) consists of a \SI{250}{\nm} gold layer applied to a commercially available silicon-nitride TEM membrane by argon plasma sputtering. Transparency for high-energy electrons was ensured by placing the \tate{} thin film above a hole with a diameter of \SI{2}{\um} drilled by focus ion beam milling. Optimal thermal properties of all sample designs was achieved by placing the samples in the TEM with the gold-side facing the laser illumination.
Besides investigations of structural dynamics by electron diffraction, such a sample design can also be employed to investigate magnetization dynamics via x-ray microscopy at up to \SI{50}{\MHz} rates \cite{zayko2021,gerlinger2023}.

\section{Thermal diffusion in the different sample support structures}
\label{sec:ThermalCharacterization}

\begin{figure*}[htbp]
\centering
\includegraphics[scale=1]{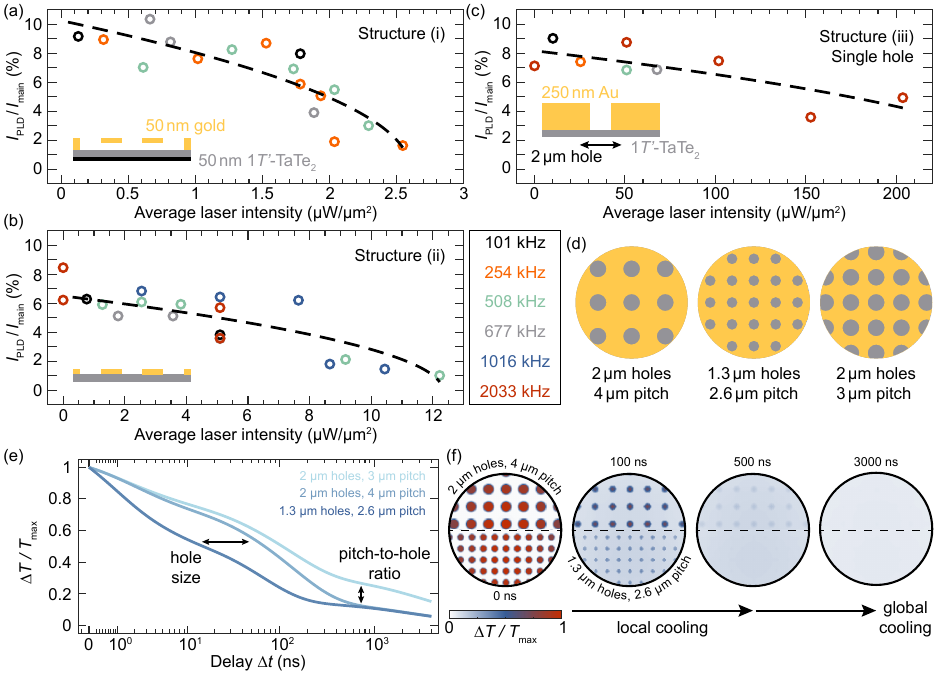}
\caption{
\textbf{Thermal characterization of sample support structures, and heat diffusion simulations.}
(a)-(c) Details on the thermal characterizations of the different sample supports described in the main text. The intensity of the $(3\times3)$ PLD satellite spots normalized by the intensity of the main lattice reflections for different fluences and repetition rates is a measure of cumulative sample heating. The average temperature increase is more pronounced in support structure (i) than in structure (ii), and suppressed most efficiently in support structure (iii). The dashed black curve corresponds to the fit result following equation~\ref{eq:ThermalFit}.
(d) Temperature evolution in the central hole for the geometries depicted in (f). The early-stage dynamics are governed by the hole diameter, where smaller values lead to a faster initial relaxation. The temperature evolution at later times is determined by the relative pitch of the gold grid, i.e., the hole-diameter to hole-distance ratio. A larger available heat bath and the reflection of the incident excitation enable a faster dissipation.
(e) Temperature evolution after the spatially structured excitation, illustrating the transition from local thermal relaxation, given by dynamics within the individual holes, to global heat dissipation determined by the overall input of heat and the hole-to-hole distance.
(f) Schematic sample geometry for the thermal transport simulations. We assume a spatially structured thermal diffusivity, i.e., holes possessing the thermal properties of \tate{} (grey) within a gold film (gold). Furthermore, the aperture array structures the optical excitation profile.
}
\label{fig:ThermalDetails}
\end{figure*}

\subsection{Thermal transport simulations}
In addition to the experimental measurements, we evaluate the suitability of the different sample designs for high-repetition-rate nano-UED by means of thermal transport simulations [cf. Fig.~\ref{fig:ThermalDetails} and Fig.~3 in the main text]. The diameter of the excitation profile exceeds the thickness of the sample supports by orders of magnitude, such that we assume the overall thermal relaxation to predominantly unfold laterally. The details of the thermal relaxation depends on a number of influences, including the different materials and interfaces between them, temperature-dependent material parameters, radiation losses and dimensionality \cite{lin1967}. Nevertheless, solving the heat diffusion equation in two dimensions quantitatively captures the behaviour of the sample support structures investigated experimentally in terms of the highest possible repetition rates in ultrafast experiments. The model reproduces the improvement from nano- to microsecond cooling times achieved by a spatially structured excitation profile and the addition of enhanced thermal dissipation provided by the gold reflection layers.

Generally, the temporal evolution of a temperature profile $T(x,y,t)$ in a two-dimensional coordinate system $(x,y)$ at time $t$ follows
\begin{align}
    \frac{\partial T}{\partial t} = \frac{1}{\rho(x,y)C_p(x,y)}\,\nabla\left[\lambda(x,y)\nabla T(x,y)\right] ,
    \label{eq:heat}
\end{align}
where $\lambda$ is the thermal conductivity, and $\rho$ and $C_p$ describe the mass density and the heat capacity, respectively. Corresponding values for \tate{} and gold are listed in Table \ref{tab:param}.

For a single material and a simple, gaussian-shaped initial temperature profile,
\begin{align}
T(x,y,0)= T_\text{max}\,\text{exp}\left[-\frac{x^2+y^2}{2\sigma^2}\right]\,,
\end{align}
equation~\ref{eq:heat} simplifies to
\begin{align}
\frac{\partial T}{\partial t}= \frac{\lambda}{\rho C_p}\,\nabla^2 T(x,y)\,,
\end{align}
and can be solved with the heat kernel \cite{rother2017}
\begin{align}
    H\left(x',y',t\right) = \frac{1}{4\pi\alpha t}\,\text{exp}\left[\frac{-\left(x'^2+y'^2\right)}{4\alpha t}\right]\, ,
    \label{eq:heat_kernel}
\end{align}
such that
\begin{align}
    T\left(x,y,t\right) = T_\text{max}\frac{2\sigma^2}{4\alpha t+2\sigma^2}\,\text{exp}\left[\frac{-\left(x^2+y^2\right)}{4\alpha t + 2\sigma^2}\right]\, .
    \label{eq:heat_solution}
\end{align}
Therein, $\alpha=\frac{\lambda}{\rho C_p}$ is a spatially-constant thermal diffusivity and $\sigma$ is the width of the initial temperature distribution, which, in the UED experiments, is given by the laser spot size. Accordingly, for large times $t$, $4\alpha t>>2\sigma^2$ and the decay time $\tau$ associated with a critical decay of the initial maximum temperature $T_\text{max}$ scales as $\sigma^2$. Qualitatively, this already explains the enhancement of the available repetition rate that is achieved by shadowing parts of the \tate{} specimen, confining the excitation to few-micrometer hotspots.

Additionally, the expected relative improvement of the thermal dissipation in the individual sample supports can be estimated based on their structural differences. The gold film thickness $l$ determines the available heat bath, the local temperature increase for a given fluence is governed by the optical reflectivities $R_\text{Au}$ and $R_\text{T}$ of gold and \tate{}, respectively, and the total absorbed energy depends on the relative gold film coverage $f$. Furthermore, cumulative heating scales linearly with the incident laser power \cite{lax1977}. Then, the maximum possible excitation intensity applied to the single-hole sample [structure (iii), $l_\text{iii}=\SI{250}{\nm}$, $f_\text{iii}=\SI{99.8}{\percent}$] exceeds the corresponding value for the gold array [structure (ii), $l_\text{ii}=\SI{50}{\nm}$, $f_\text{ii}=\SI{80.3}{\percent}$] by a factor
\begin{align}
    p = \frac{l_\text{iii}}{l_\text{ii}}\,\frac{(1-R_\text{T})\,(1-f_\text{ii})+(1-R_\text{Au})\,f_\text{ii}}{(1-R_\text{T})\,(1-f_\text{iii})+(1-R_\text{Au})\,f_\text{iii}}\, .
\end{align}
The specified parameters result in a value of $p=\num{30}$ that is in good agreement with the factor of \num{20} found experimentally [see Figure~3(c) in the main text]. In our estimation, we consider the reported refractive index $N_{\text{Au}}=n+ik$ for \num{800}\textendash\SI{}{\nm} illumination of an evaporated gold film ($n=\num{0.1244}$, $k=\num{5.04}$) \cite{olmon2012} that yields a reflectivity $R_\text{Au}=\frac{\left|1-N_{\text{Au}}\right|^2}{\left|1+N_{\text{Au}}\right|^2}$ of \SI{98}{\percent}, and a value for \tate{} of $R_\text{T}=\SI{56}{\percent}$ \cite{siddiqui2021}.

As described in the main text, accounting for spatial heterogeneities, we further integrate equation \ref{eq:heat} numerically. As initial conditions, we choose a spatially-dependent temperature profile
\begin{align}
    T(x,y,0) \propto E_{\text{max}}\,\frac{\left[1-R(x,y)\right]}{C_p(x,y)\rho(x,y)}\, ,
\end{align}
where $E_{\text{max}}$ is the incident laser energy density described as a two-dimensional Gaussian function with a full-width at half-maximum of $\SI{40}{\um}$. Furthermore, the out-of-plane structure of the individual sample supports is included by considering effective materials properties. We consider effective values for the local heat capacity
\begin{align}
    C_p(x,y)=C_{p,\text{\tate{}}}+\frac{l_{\text{Au}}}{l_{\text{\tate{}}}}\,C_{p,\text{Au}}
\end{align}
that project the available heat bath into the two-dimensional simulation plane. Both the initial excitation profile and the thermal properties of the support structures then depend on the gold film thickness, the hole diameter and the hole distance [see Fig.~\ref{fig:ThermalDetails}(f) and Fig.~3(b) in the main text]. Furthermore, the interface between the circular holes in the sample supports and the gold film is described as a linear interpolation over a quarter of the hole diameter between the properties of the bare \tate{} film in the hole regions and the surrounding combined gold/\tate{} system for all simulation parameters. This approach reduces the influence of discontinuities in the numerical simulations. Thermal resistances across the interfaces may require additional and more involved modelling beyond the main focus of this work. In particular, the simulations reproduce the enhancement of the available repetition rate found experimentally.

The simulation results are depicted in Figure~3(d) and (e) in the main text, and in Figure~\ref{fig:ThermalDetails}(d) and (e). An efficient thermal management depends on two factors: the size of the temperature hotspots, and the overall amount of absorbed heat. The former is given by the hole diameter, while the number and the distance of the holes within the optically excited region determines the latter. Accordingly, the fastest cooling time is achieved for the sample used in the measurements at \SI{2}{\MHz}, i.e., a single $\num{2}$\textendash$\mu\text{m}$ hole in an otherwise closed gold film. Our simulations yield a recovery of the initial sample temperature after around \SI{1}{\us}. In contrast, heat dissipation in a bare \tate{} thin film, possessing much lower thermal diffusivity, takes orders of magnitude longer, even without the effect of cumulative heating, i.e., repetitive laser excitation.

The dynamics in the aperture-array sample supports unfold as an intermediate between these two cases. Within the first few hundred nanoseconds, the temperature in the individual holes closely follows the case of the single-hole structure as the dissipation is given by local heat flow, and thus determined by the hole diameter. This fast thermal relaxation persists until the width of the individual temperature hotspots approaches their initial distance. From here on, $T(x,y,t)$ qualitatively evolves similar to the excitation of a bare gold film after a temperature increase with reduced magnitude. The available unexcited heat bath makes these sample designs suitable for UED experiments with repetition rates in the \SI{}{\kHz}-regime, as exemplified by the curves shown in Figure~4(b) in the main text. Furthermore, investigations of structural dynamics that can accommodate a larger degree of cumulative heating due to, e.g., a larger difference between the base and the phase transition temperature, may further extend the range of available repetition rates. Additionally, smaller hole diameters for a faster initial thermal relaxation, or larger gold array pitches to increase the available heat bath may have similar effects [Fig.~4(d)].

\begin{table}[htb]
\centering
\renewcommand{\arraystretch}{1.2}
\begin{tabularx}{\columnwidth}
{ 
  | >{\centering\arraybackslash\hsize=1\hsize}X 
  | >{\centering\arraybackslash\hsize=1\hsize}X
  | >{\centering\arraybackslash\hsize=1\hsize}X |}
\hline
    & \tate{} & Gold  \\
    \hline
    \hline
    molar mass $M$ $\left[\SI{}{\g\per\mole}\right]$ & $\num{436}$ & $\num{197}$ \\
 \hline
    density $\rho$ $\left[\SI{}{\g\per\cm^3}\right]$ & $\num{9.3}$ \cite{sorgel2006} & $\num{19.3}$ \cite{haynes2016} \\
 \hline
    thermal conductivity $\lambda$ $\left[\SI{}{\watt\per\m\per\kelvin}\right]$ & $\num{1.4}$ \cite{brixner1962} & $\num{317}$ \cite{haynes2016} \\
 \hline
 heat capacity $C_p$ & $\SI{60}{\joule\per\mole\per\kelvin}$ \cite{sorgel2006} & $\SI{129}{\joule\per\kg\per\kelvin}$ \cite{haynes2016} \\
 \hline
 thermal diffusivity $\alpha$ $\left[\SI{}{\m^2\per\s}\right]$ & $\num{1.1e-6}$ & $\num{1.27e-4}$ \\
 \hline
 reflectivity $R$ & $\SI{56}{\percent}$\cite{siddiqui2021} & $\SI{98}{\percent}$\cite{olmon2012} \\
    \hline
\end{tabularx}
\caption{Parameters used in the heat transport simulations.}
\label{tab:param}
\end{table}

\subsection{Details on the experimental characterizations}

Figure~\ref{fig:ThermalDetails}(a)\textendash(c) display the PLD spot intensities in the different support structures under variable thermal load. Measured PLD spot intensities correspond to integrated intensities, derived by summing over the detected spot profile in the individual diffractograms and over all visible second-order PLD reflections. In order to account for possible sample tilts induced by different excitation densities, we normalize the PLD signal by the intensity of all neighboring main lattice spots. For a given sample, all measurements are taken from the same sample region. Throughout the measurements, no sample damage or other permanent changes of the specimens were observed, additionally validated by prior quantitative investigations of PLD dynamics employing similar sample designs \cite{domrose2023,domrose2024}. Furthermore, no influence of the sample support on the initial PLD dynamics is found in any of these measurements, including the delay curves displayed in Figure~4 in the main text, underlining the feasibility of the structures for UED investigations of structural transitions.

In principle, the measurements displayed in Figure~3(c) and \ref{fig:ThermalDetails} average over two contributions: a thermal PLD suppression where cumulative heating reduces the spot intensity as the system moves closer to the threshold, and over times where, transiently, the PLD spot amplitude disappears completely due to the structural transition. In the ultrafast measurements, however, the sample recovers its initial low-temperature phase after a full quench already within \SI{4}{\ns}, whereas the laser pulse separation amounts to \SI{500}{\ns}. Consequently, the transient dynamics contribute around \SI{1}{\percent} to the time-averaged spot intensities. For simplicity, we therefore only consider thermal effects, and evaluate the laser-induced temperature change $\Delta T$ with a phenomenological model that includes a minimal set of parameters. We find that a square-root dependence of the PLD spot intensity $I_\text{PLD}$ on the applied laser intensity captures the observed abrupt spot suppression expected for a first-order transition close to the threshold. The cumulative heating of the individual sample supports is expected to scale linearly with the incident laser intensity $P$, i.e., $\Delta T = gP$ with a fit parameter $g$ that accounts for the thermal dissipation in the individual sample supports. Then,
\begin{align}
    I_\text{PLD}\left(\Delta T\right) = I_0\,\sqrt{1-\frac{\Delta T}{T_C-T_0}}\, ,\label{eq:ThermalFit}
\end{align}
and $I_0$ is the maximum PLD spot intensity at the estimated base temperature $T_0=\SI{120}{\kelvin}$, and the critical temperature $T_C=\SI{174}{\kelvin}$ \cite{sorgel2006}. Based on this model, we find values of $g=\SI{0.4\pm0.03}{\um^2\kelvin\per\uW}$ for structure (i), $g=\SI{0.08\pm0.006}{\um^2\kelvin\per\uW}$ for structure (ii), and $g=\SI{0.0036\pm0.0016}{\um^2\kelvin\per\uW}$ for structure (iii), representative of the improvement of the thermal dissipation found both experimentally and in the two-dimensional thermal transport simulations. The fit results are displayed in Figure~3(c) in the main text and in Figure~\ref{fig:ThermalDetails}(a)\textendash(c). A complete spot suppression corresponds to cumulative heating by $\Delta T=T_C-T_0$, observed in structure (i) and (ii). Evaluating Equation~\ref{eq:ThermalFit} at the highest laser intensity applied, we estimate a maximum temperature increase $\Delta T$ of \SI{40}{\kelvin} in structure (iii). The observed relative PLD spot suppression from \SI{8}{\percent} to \SI{5}{\percent} [Fig.~\ref{fig:ThermalDetails}(c)] corresponds to a \SI{20}{\percent} reduction of the PLD amplitude.

\bibliography{TaTe2}